# A Relational Database Model for Managing Accelerator Control System Software At Jefferson Lab[*]

S. Schaffner, T. Larrieu, JLAB, Newport News, VA 23606, USA

*Abstract*

The operations software group at the Thomas Jefferson National Accelerator Facility faces a number of challenges common to facilities managing a large body of software developed in-house. Developers include members of the software group, operators, hardware engineers and accelerator physicists. One management problem has been ensuring that all software has an identified maintainer who is still working at the lab. In some cases, locating source code for 'orphaned' software has also proven to be difficult. Other challenges include enforcing minimal standards for versioning and documentation, segregating test software from operational software, encouraging better code reuse, consolidating input/output file storage and management, and tracking software dependencies. This paper will describe a relational database model for tracking the information necessary to solve the problems above. The instantiation of that database model provides the foundation for various productivity- and consistency- enhancing tools for automated (or at least assisted) building, versioning, documenting and installation of software.

## 1 BACKGROUND

Over a number of years, the Controls Software Group at Jefferson Lab has had good success effecting formal procedures to manage software running on single board processors called input/output controllers (IOCs) that are used for the accelerator slow controls [1]. The procedures include use of the Concurrent Versioning System (CVS) to track software changes, and use custom utility scripts to standardize the management of the software and make policy-adherence less burdensome for software developers. One outcome of the system has been a decrease in accelerator downtime attributable to control software errors. The system simplified tasks such as performing system software upgrades and even Y2K auditing, however, the tools mentioned above were not enough on their own to address issues such as tracking application ownership and interdependencies.

A relational database was integrated into the system to provide the missing functionality. With information stored in the database it became possible to manage the more complex aspects of IOC configuration.

## 2 CONTROL SYSTEM UNIX ENVIRONMENT

Given the success managing IOC applications, a team was formed to develop the Control System UNIX Environment (CSUE) to similarly improve management of software running on UNIX platforms. One goal of the CSUE team is to develop and implement policies and procedures for managing operational software to ensure that all operational software is registered with an identified maintainer, and that interdependencies are documented. A second goal is to ensure that installation and modification of operational code is performed in a manner that minimizes disruption of control system operations. The system must support the segmented structure of the Jefferson Lab control system which is divided into subdomains (referred to locally as "fiefdoms" [2]), each of which has its own subnet, file and application servers, and control functions, and must typically have its own installed replica of any operationally-required software.

Several other objectives include allowing for consistent management of source code across architectures and operating systems, producing formal data file management policies, and stipulating what documentation is required before an application gets installed. As with the IOC software, scripts are used to implement the policies where ever possible. For example, all applications must be installed via a script; this script checks to make sure that any software listed as a prerequisite is already installed on the target system. Similarly, the application deletion script first checks to make sure no other applications on the target server have a dependency on the application to be deleted.

## 3   CSUE DATABASE MODEL

In order to enforce integrity rules as described in the previous section, the utility scripts need access to information about the state of servers and software across all fiefdoms. All of the information necessary for the scripts is stored in an Oracle relational database.

---

[*] Supported by DOE Contract #DE-AC05-84ER40150

The CSUE database models information in four broad categories: Developers (Staff), Applications, Systems (Servers), and Data (FileIO). The relationships among these categories are shown in Figure 1 below:

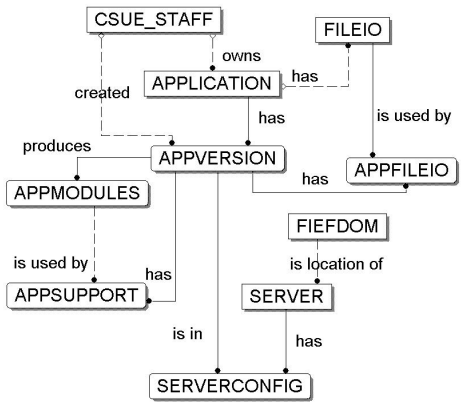

Figure 1. CSUE Database Entity-Relationship Diagram (collapsed). The child end of the relationship is denoted by the dot. Relationship phrases are read from parent to child. Solid lines denote primary key/foreign key relations. Non-identifying relations are shown with dashed lines

### 3.1 Staff

Staff information is maintained as a link to an external database that is kept in sync with Human Resource Department information. In this manner up-to-date staff information is always available, be it the most current email address for a developer, or the fact that a staff member has departed employment at Jefferson Lab.

### 3.2 Applications

In CSUE, applications are sets of source code managed as a unit. Applications may consist of mixtures of code types (e.g., C/C++, Perl, TCL, etc.), both compiled and interpreted. Functionally, an application can be code to perform an operational task or code to be used by other applications (e.g., a C/C++ library, Perl module, Tcl packages). An application may be subdivided into smaller applications, each managed separately.

Classes of applications may be defined where the details of application management differ. CSUE currently supports 2 classes of applications. The first class consists of applications whose source code is managed directly (locally written or customized applications). The second class is licensed or public-domain software (such as the Perl or Tcl interpreters) where the source is not locally modified. Different management policies apply to these types of software, but from the standpoint of the relational database model, the same set of tables can be used to store information about the software.

The Application category in the CSUE database uses several tables. The APPLICATION table stores information that is version-independent. An application owner (taken from the CSUE_STAFF table) is associated with an application. The portion of the path to the base of the source directory is also stored.

Information specific to each version is stored in the APPVERSION table. For example, the path component to the version source relative to the base source path is an attribute of this table. The relationship between the CSUE_STAFF table and the APPVERSION table is used to track the creator of the version since a team member rather than the application owner can create versions.

The products of an application are tracked in the APPMODULE table. In CSUE terms, a module is any product that will be executed or used as support by another application. Examples of products are a script, a header file, a library, or a compiled binary. Since the products of an application can vary from version to version, the APPMODULES table is related to the APPVERSION table rather than directly to the APPLICATION table.

Entire applications or portions of an application can be referenced by other applications. An example is a shared header file for compiled applications or a script application that makes an internal reference to another script from a different application. The APPSUPPORT table tracks the specific modules used by an application version from another application version (called the supporting application). The database makes it easy to determine the interdependencies between programs so that when one program is upgraded it is possible to determine in advance what other programs might be adversely affected.

### 3.3 Servers

In CSUE terms, a server is a host for executables. A server is a Unix machine, running a version of HP-UX, Linux, or Solaris on PA-RISC, Intel, or Sparc hardware. A server resides in a particular fiefdom [2] and has installed on it a subset of the applications tracked in CSUE. For any server it is possible to determine when a particular application version was in beta testing, when it was in production, when it became a rollback, and when it became obsolete.

### 3.4 File I/O

Any files produced by or used by software other than the source and its makefiles, are considered file I/O. Rather than attempt to keep track of all the files produced or used by applications, classes of file I/O are defined (in the FILEIO table) and a fixed set of file classes are defined for each application class. Versioning of file I/O is independent of application versioning because file formats, and very often documentation, change more slowly than the source code. The version of file I/O being used by a particular version of an application is tracked in the APPFILEIO table.

One major file I/O issue is management of data files: where to store data files, how long to retain them in primary storage, whether to delete old files or move them to secondary storage, what to do when a file format changes, etc. are left up to individual software developers or software users to manage. A goal of CSUE is to formalize the management of file I/O. The FileIO portion of the CSUE database provides a central store of information about the location, version, and storage policy for files produced by all applications. This information can be used to automate archival processes for file I/O. The developer is still responsible for managing the files belonging to a certain class, however, in the absence of specific instructions from a developer, all file I/O classes have a default storage/archival policy that will be applied to help the system administration team manage disk usage more effectively.

## 4 TOOLS

With the database infrastructure in place, emphasis is being placed on the refinement of tools to aid in-house development. Given the large pool of staff writing operational software, scripts, utilities, etc. it is important that capturing information about dependencies and requirements be as automatic as possible. Requiring software writers to register their applications after-the-fact using some foreign interface such as a web-form seemed likely to ensure that such registration would often not occur. So, emulating the methods for developing IOC applications [1], the team decided to develop scripts for CSUE developers to use to manage their applications. While performing useful tasks such as laying out skeleton directories for a new application, or installing a finished package, the tools capture and store critical information in the back-end database.

The tools fall into 3 categories: creation and version management tools, utility tools, and information tools. The tools interact not only with the Oracle database, but also with the file system and with the CVS repository. The tools currently available allow creating applications, specifying modules, adding support, and creating new versions. The installation tools are to be developed next. After that, work will be done to develop tools for managing the file I/O. A web-based interface to browse or search all the documentation belonging to registered applications is also in the works.

## 5 CURRENT STATUS AND FUTURE PLANS

As of this writing, the database infrastructure has been largely finalized. Nearly a dozen new applications under development have been registered in the database (including the development tools themselves) and an approximately equal number of existing applications have been imported and registered. The next challenge will be to import all of the existing software, scripts, and utilities, numbering in the hundreds, often without identified maintainers, into the CSUE structure. It will require significant effort to identify this software and import it since internal code references to other software and to files will have to be modified to reference standard CSUE locations and paths. In addition, developers will be forced to think about such issues as file management and proper documentation. This may prove to be as difficult as actually developing the database and the tools.